\documentclass[twocolumn,pra,aps,showpacs]{revtex4}

\usepackage{graphicx}
\usepackage{dcolumn}
\usepackage{amsmath}

\begin{document}

\title{Vortex lattice formation in a rotating Bose-Einstein condensate} 
\author{Makoto Tsubota$^1$}
\author{Kenichi Kasamatsu$^1$}
\author{Masahito Ueda$^2$}
\address{%
$^{1}$
Department of Physics,
Osaka City University, Sumiyoshi-Ku, Osaka 558-8585, Japan \\
$^{2}$ Department of Physics, Tokyo Institute of Technology,  
Meguro-ku, Tokyo 152-8551, Japan
}%

\date{\today}

\begin{abstract}
We study the dynamics of vortex lattice formation of a rotating 
trapped Bose-Einstein condensate by 
numerically solving the two-dimensional Gross-Pitaevskii 
equation, and find that the condensate undergoes elliptic deformation, 
followed by unstable surface-mode excitations before forming 
a quantized vortex lattice. The origin of the peculiar surface-mode 
excitations is identified to be phase fluctuations 
at the low-density surface regime. The obtained dependence of 
a distortion parameter on time and that on the driving frequency 
agree with the recent experiments by Madison {\it et al.} 
[Phys. Rev. Lett. {\bf 86}, 4443 (2001)]. 

\end{abstract}

\maketitle

Quantized vortices have long been studied in superfluid $^4$He
as the topological defects characteristic of 
superfluidity \cite{Donnelly,Pismen}.
However, the relatively high density and strong repulsive interaction
complicate the theoretical treatments of the Bose-Einstein condensed liquid,
and the healing length of the atomic scale makes
the experimental visualization of the quantized vortices difficult.
The recent achievement of Bose-Einstein condensation in
trapped alkali-metal atomic gases at ultra low temperatures has stimulated
intense experimental and theoretical activity.
The atomic Bose-Einstein condensates(BECs) have the weak 
interaction because they are dilute gases, thus being free of the 
above difficulties that superfluid $^4$He is subject to.
Quantized vortices in the atomic BECs have recently been created 
experimentally by Matthews {\it et al.} \cite{JILA}, 
Madison {\it et al.} \cite{ENS1} 
and Abo-Shaeer {\it et al.} \cite{MIT}.

By rotating an asymmetric trapping potential, Madison {\it et al.} at ENS 
succeeded in forming a quantized vortex in $^{87}$Rb BEC 
for a stirring frequency that exceeds 
a critical value \cite{ENS1}.
Vortex lattices were obtained for higher frequencies.
The ENS group subsequently observed that vortex nucleation occurs 
via a dynamical instability of the condensate \cite{ENS2}.
For a given modulation amplitude and stirring frequency, 
the steady state of the condensate
was distorted to an elliptic cloud, stationary in the rotating frame, as
predicted by Recati et al. \cite{Recati}.
An intrinsic dynamical instability \cite{Sinha} 
of the steady state transformed
the elliptic state into a more axisymmetric state with vortices.
However, the origin of that instability and how it leads to the 
formation of vortex lattices remains to be investigated

The ENS group found that the minimum rotation 
frequency $\Omega_{\rm nuc}$ at which
one vortex appears is 0.65$\omega_{\perp}$, 
where $\omega_{\perp}$ is the transverse oscillation
frequency of the cigar-shaped trapping potential, 
independent of the number of atoms or the
longitudinal frequency $\omega_z$ \cite{ENS3}.
There is a discrepancy between the observations and the theoretical
considerations based on the stationary solution of the Gross-Pitaevskii
equation(GPE) \cite{Isoshima,cri};
$\Omega_{\rm nuc}$ is significantly larger than its equilibrium estimates.

The present paper addresses these issues by numerically solving 
the GPE that governs the time evolution of the order parameter 
$\psi({\bf r},t)$:
\begin{eqnarray}
(i - \gamma)\hbar \frac{\partial \psi}{\partial t}
= \Bigl[ -\frac{\hbar^2}{2m}\nabla^2+V_{\rm tr}+ g|\psi|^2  
-\mu-\Omega L_z \Bigr] \psi.
\label{GPE}
\end{eqnarray}
Here $g=4\pi \hbar^2 a/m$ is the coupling constant, proportional to the
$^{87}$Rb
scattering length $a\approx$5.77 nm.
The high anisotropy of the cigar-shaped potential used in the ENS
experiments ($\omega_{\bot}/\omega_z \sim 14$) may permit 
the two-dimensional analysis.
We thus focus on the two-dimensional dynamics of Eq. (\ref{GPE}) 
by assuming the trapping potential
\begin{equation}
V_{\rm tr} ({\bf r})= \frac12 m \omega_{\perp} ^2 
\{ (1+\epsilon_x)x^2+(1+\epsilon_y)y^2 \},
\label{potential}
\end{equation}
where $\omega_{\perp}= 2\pi \times 219$ Hz, 
and the parameters $\epsilon_x=0.03$ and
$\epsilon_y=0.09$
describe small deviations of the trap from the axisymmetry, corresponding to
the ENS experiments \cite{ENS1}.
The centrifugal term $-\Omega L_z=i \hbar \Omega (x\partial_y -
y\partial_x)$
appears in a system rotating about the $z$ axis at a frequency $\Omega$.
An important characteristic parameter of the two-dimensional system is
$C=8\pi Na/L$, with the total number $N$ of the condensate
atoms and
$L$ the size of the system along the $z$ axis.

\begin{figure*}[ht]
\includegraphics[height=0.33\textheight]{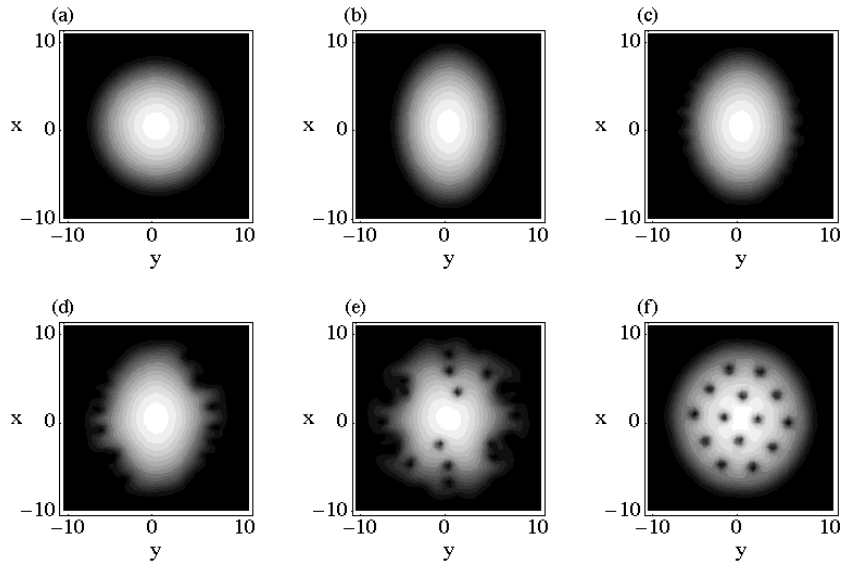}%
\caption{Time development of the condensate density $|\psi|^2$ 
after the trapping potential
begins to rotate suddenly with $\Omega=0.7\omega_{\perp}$.
The time is $t=$0 msec (a), 21 msec (b), 107 msec (c), 114 msec (d), 123
msec (e), and 262 msec (f). The unit for length is 
$a_{\rm HO} = \sqrt{\hbar/2 m \omega_{\perp}}=0.512 
 \mu$m and the period of the trap $4.57$ msec.}
\label{vorden}
\end{figure*}

The term with $\gamma$ in Eq.\ (\ref{GPE}) introduces the dissipation.
Although the detailed mechanism of the dissipation is yet to be understood, 
we include this term in the GPE because of the following reasons.
First, collective damped oscillations of the condensate have been observed
by Jin {\it et al.} and Mewes {\it et al.} \cite{MIT2}, 
which shows the presence of some dissipative mechanisms and
is consistent with the solution of the GPE with the 
dissipative term \cite{Choi}.
Second, even if the trapping potential is rotated fast enough, vortex
lattices will never be formed without dissipation, 
because vortex lattice correspond to local
minimum of the total energy in the configuration space \cite{rota}.
In other words, the observation of the 
vortex lattices \cite{ENS1} implies the
presence of dissipation.
Following Ref. \cite{Choi}, we use $\gamma =0.03$ throughout this work.

In this paper, we focus on two kinds of responses of the condensate, namely,
the response to the sudden turn on of the rotation of the potential, and
that to the slow turn on. The numerical calculations are performed with the
implicit Crank-Nicolson method. The chemical potential 
$\mu$ is adjusted at all 
times so as to preserve the number of condensate atoms.

We first prepare an equilibrium condensate
with $C=1400$ trapped in a stationary potential.
Figure 1 shows the typical dynamics of the 
condensate density $|\psi({\bf r},t)|^2$
after the
potential begins to rotate suddenly with $\Omega = 0.7\omega_{\bot}$ \cite{HP}.
The condensate is elongated along the $x$ axis because of the small anisotropy
of $V_{\rm tr}$ [Eq.\ (\ref{potential})], and the elliptic cloud oscillates.
Then, the boundary surface of the condensate becomes unstable,
exciting the surface waves, which propagate along the surface.
The excitations are likely to occur on the surface 
whose curvature is low, i.e.,
parallel to the longer axis of the ellipse.
The ripples on the surface develop into the vortex cores around
which superflow circulates.
Subject to the dissipative vortex dynamics, 
some vortices enter the condensate,
forming a vortex lattice. As the vortex lattice is being formed, 
the axial symmetry of the condensate is recovered by transferring 
angular momentum into quantized vortices.

This peculiar dynamics is understood by investigating 
the phase of $\psi({\bf r},t)$ as shown in Fig. 2. 
There are some lines where the phase changes discontinuously 
from black to white, which corresponds to the branch cuts 
between the phases 0 and 2$\pi$. 
Their ends represent phase defects, i.e. vortices.
As soon as the rotation starts,
some defects begin to enter.
When the defects are on the outskirts of the condensate
where the amplitude $|\psi({\bf r},t)|$ is
almost negligible, they neither contribute to the energy nor the angular
momentum of the system.
These defects come into the boundary surface of the condensate within which
the amplitude grows up. Then the defects compete with each other and 
induce the above surface waves due to interference.
There the selection of the defects starts, because their further invasion
into the condensate costs the energy and the angular momentum.
As is well known in the study of rotating superfluid helium \cite{rota},
the rotating drive pulls vortices
into the rotation axis, while repulsive interaction tends to push them
apart; this competition yields a vortex lattice whose vortex density 
depends on the rotation frequency.
In our case,  some vortices enter the 
condensate and form a lattice dependent on
$\Omega$, while excessive vortices are repelled and escape to the outside.
Remarkably, the phase profile of Fig. 2(b) reveals that the repelled vortices
also form a lattice on the outskirts of the condensate. Since they cannot
be seen in the corresponding density profile of Fig. 1(f), they may be called
``ghost" vortices.

\begin{figure}[ht]
\includegraphics[height=0.17\textheight]{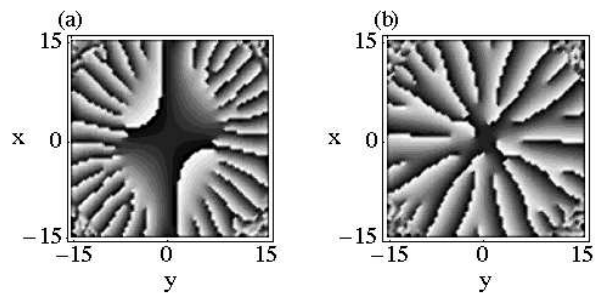}%
\caption{Phase profile of $\psi$: (a) and (b) 
corresponds to Fig. 1 (c) and 1 (f). The
value of the phase varies continuously from $0$ (black) to $2 \pi$ (white). 
The unit for length is the same as that of Fig. 1}
\label{phase}
\end{figure}

The distortion of the condensate to an elliptic 
cloud was theoretically studied
by Recati {\it et al.} \cite{Recati}.
Assuming the quadrupolar velocity field 
${\bf v}({\bf r}) = \alpha \nabla (xy)$,
they obtained the distortion parameter
$\alpha=\Omega (R_x^2 - R_y^2)/(R_x^2 + R_y^2)$ in the steady states as a
function of $\Omega$, where $R_x$ and $R_y$ are the sizes of the
condensate along each direction.
Madison {\it et al.} observed that, after the rotation 
of $\Omega = 0.7\omega_{\bot}$
starts suddenly,  $\alpha$ oscillates during a few hundred
milliseconds and then falls abruptly to a value below 0.1 when vortices
enter the condensate from its boundary surface \cite{ENS2}.
Figure 3 shows the oscillation of $\alpha$ and the increase of the angular
momentum $\ell_z/\hbar$ per atom in our dynamics of Fig. 1.
This figure closely resembles Fig. 3 of Ref. \cite{ENS2}, 
and our scenario is thus 
consistent with the experimental results.

\begin{figure}[bp]
\includegraphics[height=0.19\textheight]{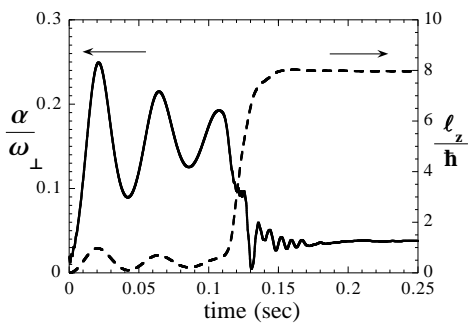}%
\caption{Time evolution of the distortion parameter $\alpha$ (solid line) 
and the angular momentum $\ell_{z}/\hbar$ per atom (dashed line) 
corresponding to the dynamics of Fig. 1.}
\label{disto}
\end{figure}

The dispersion relation of the surface waves \cite{Khawaja} of a rotating
condensate is obtained as follows.
The substitution of the Madelung transformation $\psi=|\psi|e^{i \theta}$
into Eq. (\ref{GPE}) yields
\begin{subequations}
\begin{eqnarray}
\frac{\partial |\psi|^2}{\partial t}+\nabla \cdot \left[ |\psi|^2 ({\bf v}-
\Omega \times {\bf r})  \right] =0, \\
m\frac{\partial {\bf v}}{\partial t}=-\nabla 
\biggl\{ \delta\mu +\frac{m}{2} \left( {\bf v}-
\Omega \times {\bf r} \right)^2-\frac{m}{2} \Omega^2 r^2 \biggr\},
\end{eqnarray}
\end{subequations}
where ${\bf v}=(\hbar/m)\nabla \theta$ and $\delta \mu =V_{\rm tr} 
+ g|\psi|^2 - \frac{\hbar^2}{2m|\psi|}\nabla^2 |\psi| - \mu$.
By linearizing Eq. (3), we obtain the dispersion relation
of the surface waves $\omega^2=R(\omega_{\bot}^2-\Omega^2)k$ with the
condensate size $R$.
The dependence of $\omega$ on $R$ shows that the surface with lower curvature
is excited more easily, which is clearly seen in Fig. 1.
The group velocity $d\omega/dk= \sqrt{R(\omega_{\bot}^2-\Omega^2)}/(2\sqrt{k})$
agrees with the propagation velocity of the surface wave in our simulation.
As discussed by Al Khawaja {\it et al.} \cite{Khawaja}, the surface waves are
connected with the low energy excitations studied by Stringari 
and Dalfovo {\it et al.} \cite{Stringari}, and
Isoshima and Machida \cite{Isoshima}.

We have examined the critical frequency at which
vortices can enter with $C=420$ corresponding to the experimental
condition \cite{ENS1}, and found that only when $\Omega$ is larger than 
$\Omega_{c1} \simeq 0.57\omega_{\perp}$ 
can vortices enter the condensate and form a lattice.
This critical frequency is closer to the observed value $\Omega_{\rm nuc} =
0.65\omega_{\perp}$,
than the values obtained in previous literature.
The critical frequencies have been studied from thermodynamic or stability
arguments \cite{Isoshima,cri,3D}, but they are generally much 
smaller than the observed value.
It should be noted that these critical frequencies only give the necessary
condition which enables a vortex to exist stably at the center of the trap.
Actually, vortices should be nucleated at the boundaries and come into the
condensate; the condition of $\Omega$, which realizes such nonlinear dynamics,
may be generally different from that obtained from the stability arguments.
Isoshima and Machida examined the local instability of nonvortex states
toward vortex states within the Bogoliubov theory \cite{Isoshima}.
They note that this local instability could give higher critical frequencies 
than the global instability, which comes from comparing their total energy,
for example, about $0.6\omega_{\perp}$ 
for the ENS experimental condition \cite{Machida}.
The stability analysis cannot answer what happens when 
the system goes over the linear region.
Our paper reveals the importance of the 
nonlinear dynamics beyond the stability analysis.
The detailed studies of the dependence of the number of vortices on $\Omega$,
that of $\Omega_{c1}$ on the number of condensate atoms 
\cite{Isoshima} will be reported elsewhere.

Finally, we study the evolution of the condensate in a
potential rotating with a time-dependent frequency $\Omega(t)$.
The experiments were made by Madison {\it et al.} \cite{ENS2}.
The condensate at every moment was in a stationary state with an elliptic
cloud obtained by Recati {\it et al.} under the Thomas-Fermi limit \cite{Recati}, 
since the time dependence was slow enough.
One of the important observations was that, for an ascending ramp, the
condensate follows branch I of the nonvortex state 
until $\Omega_c \simeq 0.75 \omega_{\perp}$.
Beyond this critical frequency, the condensate changes
from branch I to the more axisymmetric state, nucleating vortices.
This critical frequency agrees with the frequency at which branch I becomes
dynamically unstable \cite{Sinha}; however, the detailed nature of the
bifurcation has remained to be clarified.
We study the response of the condensate of $C=420$ for the ascending frequency
$\Omega(t)=ct$ with $c=2\pi \times 400$ Hz/s; they are about the same values
as those used in the ENS experiment.
The obtained results are consistent with the experimental ones, as shown in
Fig. 4. As $\Omega$ increases, the condensate becomes gradually 
elliptic following branch I,
when the nucleated ``ghost" vortices come to the surface of the condensate.
When $\Omega$ is raised to the 
critical value $\Omega_{c2} \simeq 0.75 \omega_{\perp}$,
the vortices jostle with each other and excite the surface waves, reflecting
the dynamical instability.
Then, similar to the first set of calculations, 
some vortices enter the condensate, form
a lattice and increase the angular momentum, 
while the condensate recovers the axisymmetry
apart from branch I. The small discrepancy 
of the trace of our calculation from
branch I in Fig. 4 may be attributed to a deviation from the
Thomas-Fermi limit; the similar discrepancy is reported by
the Madison {\it et al.} too \cite{ENS2}.

\begin{figure}[ht]
\includegraphics[height=0.19\textheight]{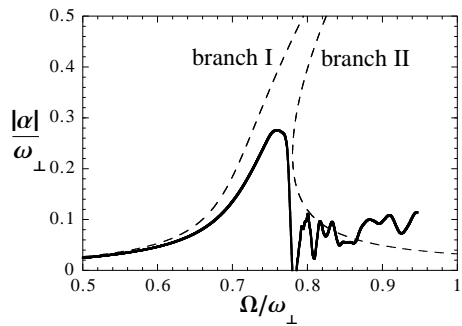}%
\caption{Development of $\alpha$  for  the ascending frequency $\Omega(t)$.
The solid curve describes our numerical result, while the dashed curve
represents the stationary nonvortex state obtained by Recati 
{\it et al.} \cite{Recati}.
See the text for more details.
}
\end{figure}

It is necessary to compare two critical frequencies $\Omega_{c1} \simeq 0.57
\omega_{\perp}$ and
$\Omega_{c2} \simeq 0.75 \omega_{\perp}$.
What causes the difference?
In the first set of calculations where the potential begins to rotate 
abruptly with a frequency $\Omega$, the condensate first deviates 
much from the stationary
state $\alpha(\Omega)$ with the quadrupolar velocity field.
As shown in Fig. 3, it seemingly relaxes to that state through the damped
oscillation
of the large amplitude, but it changes on the way to another state with
a vortex lattice which
has lower energy than the nonvortex state $\alpha(\Omega)$.
The critical frequency 
$\Omega_{c1}$ means that the state with vortices can be
realized dynamically from the nonvortex 
state only when $\Omega > \Omega_{c1}$.
In the second calculation, as $\Omega$ increases slowly, the condensate
follows
quasistatically the nonvortex state $\alpha(\Omega)$ of branch I, even after
it becomes metastable for $\Omega > \Omega_{c1}$.
Only when $\Omega$ exceeds $\Omega_{c2}$, the condensate can change from
this state
to the vortex state, reflecting the dynamical instability \cite{Sinha}.

Recently Garc\'{i}a-Ripoll {\it et al.} and Feder {\it et al.} study this system by
three-dimensional
analysis of the GPE \cite{3D}, taking account of the bending of vortex lines
in the
cigar-shaped potential.
A bent vortex moves with the self-induced velocity proportional to its
curvature \cite{Donnelly}, which is lacking in the two-dimensional dynamics.
However, due to the high
anisotropy of the cigar-shaped potential, 
vortices are aligned almost straight
along the symmetry axis. It is therefore unlikely that the bending effect
alters the results of our work qualitatively and significantly.

In conclusion, we have studied the dynamics of vortex lattice formation of
a rotating trapped BEC by numerically solving the two-dimensional GPE, and
obtained the following physical picture.
When the trapping potential begins to rotate 
at a sufficiently fast frequency,
the condensate is distorted to an elliptic shape and oscillates.
Then, its boundary surface becomes unstable, exciting surface waves.
The origin of these ripples
is identified to be the interference of violent phase fluctuations 
that occur on the outskirts of the condensate; 
and some of the surface ripples  develop
into the vortex cores which then form a vortex lattice.
The critical frequency for this process is found to be $\Omega_{c1} \simeq
0.57 \omega_{\perp}$ for the condition at the ENS group.
On the other hand, if the rotation frequency is raised slowly from zero, the
condensate
becomes elliptic and follows the nonvortex elliptic state with the
quadrupolar flow \cite{Recati}.
When $\Omega$ exceeds $\Omega_{c2} \simeq 0.75 \omega_{\perp}$, 
the condensate deviates
from the state,
nucleating vortices via the dynamical instability \cite{Sinha}.
The whole dynamics is concerned with the ``ghost" vortices, which are the
phase defects
outside the condensate.
The obtained results are consistent with the experimental results of the ENS
group \cite{ENS1,ENS2}.

The authors thank K. Machida for useful discussions. M.U. acknowledges support 
by a Grant-in-Aid for Scientific Research
(Grant No.~11216204) by the Ministry of Education, Science, Sports,
and Culture of Japan, and by the Toray Science Foundation.


\begin{references}
\bibitem{Donnelly}
R. J. Donnelly, {\it Quantized Vortices in Helium II}
(Cambridge University Press, Cambridge, 1991).
\bibitem{Pismen}
L. M. Pismen, {\it Vortices in Nonlinear Fields}
(Oxford University Press, Oxford, 1999).
\bibitem{JILA} M. R. Matthews {\it et al.}, Phys. Rev. Lett. 
{\bf 83}, 2498 (1999).
\bibitem{ENS1} K. W. Madison {\it et al.}, Phys. Rev. Lett. 
{\bf 84}, 806 (2000).
\bibitem{MIT} J. R. Abo-Shaeer {\it et al.}, Science
{\bf 292}, 476 (2001).
\bibitem{ENS2} K. W. Madison {\it et al.},  Phys. Rev. Lett. 
{\bf 86}, 4443 (2001).
\bibitem{Recati} A. Recati, F. Zambelli, and S. Stringari,
Phys. Rev. Lett. {\bf 86}, 377 (2001)
\bibitem{Sinha} S. Sinha and Y. Castin, Phys. Rev. Lett. {\bf 87},
190402 (2001).
\bibitem{ENS3} F. Chevy {\it et al.}, Phys. Rev. Lett. {\bf 85},
2223 (2000).
\bibitem{Isoshima} T. Isoshima and K. Machida, J. Phys. Soc. Jpn. {\bf 68},
 487 (1999); Phys. Rev. A {\bf 60}, 3313 (1999).
\bibitem{cri}G. Baym {\it et al.}, Phys. Rev. Lett. {\bf 76}, 6 (1996);
F. Dalfovo {\it et al.}, Phys. Rev. A {\bf 53}, 2477 (1996); S. Sinha,
{\it ibid.} {\bf 55}, 4325 (1997); E. Lundth {\it et al.}, {\it ibid.}
{\bf 55}, 2126 (1997); A. A. Svidzinsky {\it et al.}, Phys. Rev. Lett.
{\bf 84}, 5919 (2000); Y. Castin {\it et al.}, Eur. Phys. J. D {\bf 7}, 399 
(1999); F. Dalfovo {\it et al.}, Phys. Rev. A {\bf 63}, 011601(R) (2001).
\bibitem{MIT2} D. S. Jin {\it et al.}, Phys. Rev. Lett.
{\bf 77}, 420 (1996); M. -O. Mewes {\it et al.}, {\it ibid.}
{\bf 77}, 988 (1996).
\bibitem{Choi} S. Choi, S. A. Morgan, and K. Burnett, Phys. Rev. A
{\bf 57}, 4057 (1998).
\bibitem{rota} L. J. Campbell and R. M. Ziff, Phys. Rev. B {\bf 20},
1886 (1979); M. Tsubota and H. Yoneda, J. Low Temp. Phys. {\bf 101},
815 (1995).
\bibitem{HP} You can see the animation of this dynamics in
http://matter.sci.osaka-cu.ac.jp/bsr/vortexex-e.html.
\bibitem{Khawaja}U. Al Khawaja, C. J. Pethick, and H. Smith,
Phys. Rev. A {\bf 60}, 1507 (1999).
\bibitem{Stringari}S. Stringari, Phys. Rev. Lett. {\bf 77}, 2360 (1996);
F. Dalfovo {\it et al.}, Phys. Rev. A {\bf 56}, 3840 (1997).
\bibitem{3D}J. J. Garc\'{i}a-Ripoll {\it et al.}, Phys. Rev. A {\bf 63}, 
041603 (2001); D. L. Feder {\it et al.}, Phys. Rev. Lett. {\bf 86}, 564 (2001).
\bibitem{Machida} K. Machida (private communication).

\end{references}
\end{document}